\documentclass[journal=langd5, manuscript=article, layout=twocolumn]{achemso}

\usepackage[version=3]{mhchem} % Formula subscripts using \ce{}
\usepackage[T1]{fontenc}       % Use modern font encodings
\usepackage{amsmath, amssymb, amsfonts}
\usepackage{graphicx, color, gensymb}

\SectionNumbersOn
\graphicspath{ {figures/} }
\author{V. S. Akella}
\affiliation{Collective Interactions Unit, OIST Graduate University, Okinawa, Japan 904-0495}
\author{D. K. Singh}
\affiliation{Collective Interactions Unit, OIST Graduate University, Okinawa, Japan 904-0495}
\author{S. Mandre}
\affiliation{School of Engineering, Brown University, 182 Hope Street, Providence, RI 02906, USA}
\email{shreyas_mandre@brown.edu}
\author{M. M. Bandi}
\affiliation{Collective Interactions Unit, OIST Graduate University, Okinawa, Japan 904-0495}
\email{bandi@oist.jp}

\title[]{Dynamics of a Camphoric Acid boat at the air-water interface} 

\abbreviations{CA, DI}

\begin{document}

% The "tocentry" environment can be used to create an entry for the
% graphical table of contents. It is given here as some journals
% require that it is printed as part of the abstract page. It will
% be automatically moved as appropriate.

% \begin{tocentry}

% Some journals require a graphical entry for the Table of Contents.
% This should be laid out ``print ready'' so that the sizing of the
% text is correct.

% Inside the \texttt{tocentry} environment, the font used is Helvetica
% 8\,pt, as required by \emph{Journal of the American Chemical
% Society}.

% The surrounding frame is 9\,cm by 3.5\,cm, which is the maximum
% permitted for  \emph{Journal of the American Chemical Society}
% graphical table of content entries. The box will not resize if the
% content is too big: instead it will overflow the edge of the box.

% This box and the associated title will always be printed on a
% separate page at the end of the document.

% \end{tocentry}

% The abstract environment will automatically gobble the contents
% if an abstract is not used by the target journal.

\begin{abstract}
We report experiments on an agarose gel tablet loaded with camphoric acid (c-boat) set into self-motion by interfacial tension gradients at the air-water interface. We observe three distinct modes of c-boat motion: harmonic mode where the c-boat speed oscillates sinusoidally in time, a steady mode where the c-boat maintains constant speed, and a relaxation oscillation mode where the c-boat maintains near-zero speed between sudden jumps in speed and position at regular time intervals. Whereas all three modes have been separately reported before in different systems, we show they belong to a common description. Through control of the air-water surface tension with Sodium Dodecyl Sulfate (SDS), we experimentally deduce the three self-propulsive modes result from surface tension difference between Camphoric Acid (CA) and the ambient surroundings.
\end{abstract}

% Start the main part of the manuscript here.

\section{Introduction}
\label{introsec}
Studies on the self-motion of camphor at air-water interfaces have a distinguished history in the annals of science. Between first reported observations in 1686 \cite{Heyde1686} and the eventual explanation of camphor self-motion in 1869 \cite{Mensbrugghe1869} as resulting from surface tension gradients, the problem attracted the attention of some of the finest scientific minds, including Alessandro Volta \cite{Volta1787}, Giovanni Battista Venturi \cite{Venturi1797}, Jean-Baptiste Biot \cite{Biot1801} and Lord Rayeligh \cite{Rayleigh1889} among others; for a historical introduction up until 1869 please see Charles Tomlinson's excellent review \cite{Tomlinson1869}. Despite its rich history, the camphor boat system continues to remain relevant into modern times, be it in statistical mechanics within the context of active matter \cite{Ramaswamy2010}, hydrodynamic context of viscous marangoni propulsion \cite{Lauga2012}, biological context of chemomechanical transduction \cite{Nakata1997}, autonomous motion and self-assembly \cite{Ismagilov2002}, and reconfigurable actuators in soft matter physics \cite{Geryak2014}, among others.

In this article, we present an experimental study of the self-motion of agarose gel tablets loaded with camphoric acid (CA) at the air-water interface, henceforth referred to as c-boats. We identify three distinct modes of motion, namely a harmonic mode where the c-boat speed undergoes time-varying sinusoidal oscillations, a steady mode of constant c-boat motion, and a relaxation oscillation mode where the c-boat remains at rest between sudden jumps in speed and position at regular time intervals. Whereas all three modes have been separately reported in the published record \cite{Hayashima2001, Suematsu2010, Jin2012, Velev2012} in a variety of systems, we show these seemingly different self-propulsive modes arise from a common description. Through metered dosage of Sodium Dodecyl Sulfate (SDS) to control the air-water surface tension, we experimentally trace the origin of self-propulsive mode selection to CA-water surface tension difference.

\section{Experimental Methods}
\label{expsec}
\begin{figure}[ht]
\centering
\includegraphics[width=0.9\linewidth]{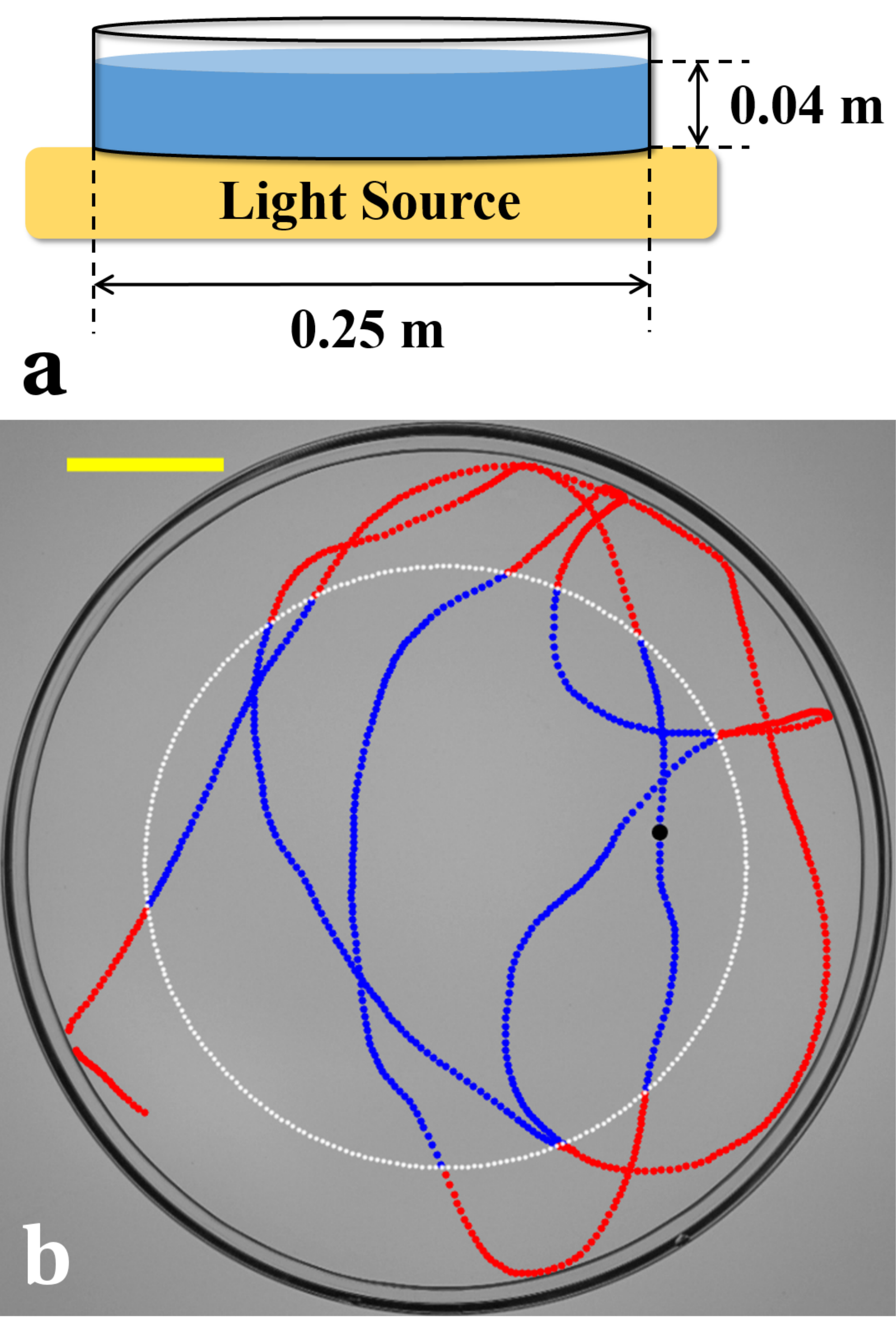}
\caption{(a) Experimental setup (side view): Glass petri dish (0.25 m diameter) filled with deionized water to 0.04 m height was placed on a light tablet. A camera recorded c-boat self-motion from above yielding images as shown in (b) where the c-boat appears as a dark disk in grey background. Only trajectories (blue) in a 0.178 m diameter circular region within the dashed white circle were analysed and the rest (red) discarded to exclude boundary effects from petri dish wall. Scale bar = 0.03 m.}
\label{fig1}
\end{figure}

Figure~\ref{fig1}a shows a schematic of the experimental setup. All experiments were performed in a glass petri dish (0.25 m in diameter) filled with de-ionized water to a height of 0.04 m. A camphoric acid tablet (c-boat) was gently introduced at the air-water interface and its self-motion was recorded with a Nikon D800E camera at 30 frames per second. The petri dish was placed atop a uniform backlit LED illumination source operating with direct current to avoid alternating current flicker interference in image processing.  The c-boat appears as a dark disk moving in a bright background in this imaging method as shown in fig.~\ref{fig1}b. The experimental images were post processed with image analysis algorithms written in-house to obtain the c-boat position and velocity as a function of time. The c-boat position and velocity information employed in the analysis was confined to a region 0.036 m away from the walls to exclude boundary effects. The portion of c-boat trajectory (red in fig.~\ref{fig1}b) lying outside the dashed white circle in fig.~\ref{fig1}b at a distance of 0.036 m from petri dish wall were excluded from the analysis. This 0.036 m exclusion distance was empirically determined from the longest radial distance over which marangoni spreading of camphoric acid was prominent (see section~\ref{resultsec}). Only blue sections of the c-boat trajectory within the inner circle bounded by white dashed line in fig.~\ref{fig1}b were used in all the analysis to follow.

The shape and size of a chemical-loaded tablet, e.g. a camphoric acid tablet, changes over the duration of the experiment as the substance undergoes dissolution or sublimation into the surrounding fluids; water and air respectively. We constructed c-boats by infusing camphoric acid in agarose gel tablets similar in spirit to the procedure of Soh et al \cite{Soh2008}, thus keeping tablet shape intact for the entire duration of the experiment. Hot agarose solution (5\% weight-to-volume) in de-ionized (DI) water (Milli-Q Integral Water Purification System with resistivity, $\rho=18.2$ M$\Omega\cdot$cm at 25$^{\circ}$C) was placed between two clean glass plates, set $10^{-3}$ m apart with aluminum spacers, to obtain gel sheets of uniform $10^{-3}$ m thickness, upon cooling. Gel tablets of $3 \times 10^{-3}$ m diameter were punched out from the sheet (Biopunch, Ted Pella Inc.). These gel tablets were introduced in a saturated solution of camphoric acid (CA) (Wako Pure Chemical Industries, Ltd., Cat. No. 036-01002) in methanol and left for 2 hours for CA to diffuse into the gel tablets. Prior to experiments, gel tablets were rinsed in DI water to precipitate CA in the gel matrix. 

The c-boat motion being governed by Marangoni force, surface tension difference between the ambient surface and CA entering the surface forms the primary parameter in this study. Since the c-boat holds a finite quantity of CA, its concentration monotonically decreases with time, thereby continually reducing the strength of the marangoni effect (see section~\ref{resultsec}). Independent experiments modifying the ambient surface tension confirmed the role of surface tension difference as the primary parameter. We varied the surface tension of the ambient interface by introducing metered dosage of Sodium Dodecyl Sulfate (SDS) (Wako Pure Chemical Industries, Ltd., Cat. No. 196-08675) following published tables \cite{Mysels1986}. Actual surface tension values were also independently confirmed with the pendant drop method on a tensiometer (OneAttension Theta tensiometer) at 25 $^{\circ}$C.

A moving c-boat leaves camphoric acid in its wake which can be visualised by introducing buoyant tracer particles that decorate the air-water interface (fig.~\ref{fig2} e-f). Hollow silica glass spheres (specific gravity 0.25 and $50 \pm 10 \times 10^{-6}$m diameter) sprinkled onto the air-water interface were used to follow the well defined comet-shaped particle-free region in the wake of a c-boat. The shape of this region provides a qualitative measure of the strength of marangoni force acting on the c-boat (see supplementary information). Experimental visualisation requires high tracer concentrations at which they do not behave passively, but instead provide a back reaction force against the spreading CA. We could not measure this back reaction force for a given tracer concentration. We therefore kept the tracer concentration fixed and studied how the tracer free, CA rich region around the moving c-boat varies with time as CA concentration changes. We were able to quantify the change in the CA rich, tracer free region as a function of time for the fixed tracer concentration.

\section{Results}
\label{resultsec}
\begin{figure*}[ht]
\centering
\includegraphics[width=0.85\linewidth]{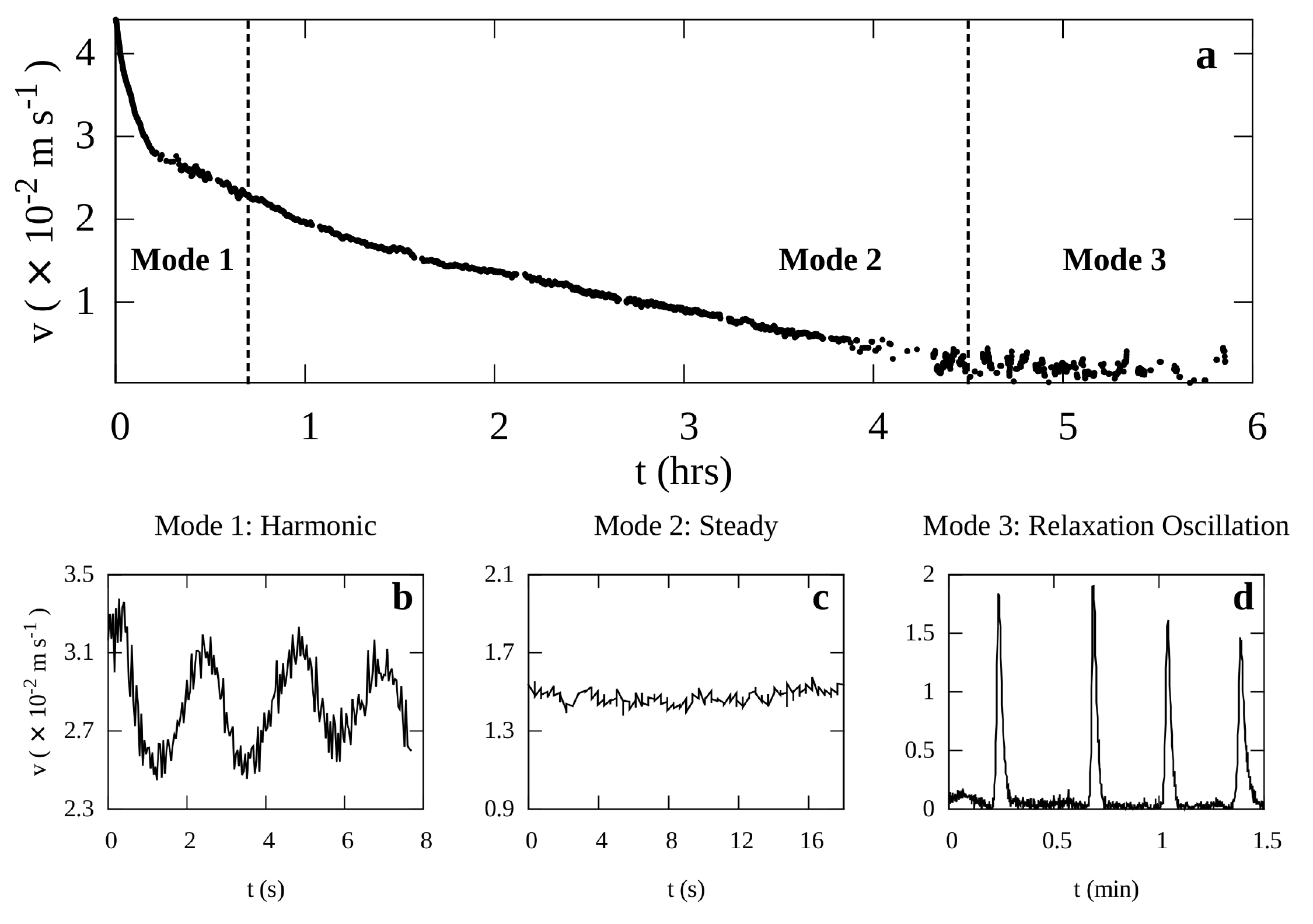}
\caption{(a) Monotonic decrease in mean c-boat speed (1 minute running average) over a 6 hour period, during which the c-boat exhibits three distinct modes of motion: (b) Harmonic mode with oscillating speed, (c) steady mode with constant speed, and (d) relaxation oscillation mode with long intervals of near-zero speed with sudden jumps at regular intervals.}
\label{fig2}
\end{figure*}

When a c-boat is placed at the air-water interface, CA spreads radially and sets up axisymmetric interfacial tension gradients around the c-boat. Ambient fluctuations spontaneously break this symmetry and sharpen the gradients along a preferential direction; as a consequence a net force acts on the c-boat and propels it. The boat's motion amplifies the asymmetry and maintains it, thereby permitting it's continued motion. Owing to continuous CA dissolution from the interface into the bulk fluid, the c-boat motion continues until it exhausts all CA molecules. Whereas dissolution does globally reduce the surface tension of water, a single boat contains  ($\sim 7 \times 10^{-6}$ kg) CA on average and is insufficient to achieve an appreciable reduction in interfacial tension {\it via} dissolution; surface tension of CA saturated water ($\sim 8 \times 10^{-3}$ kg$\cdot$l$^{-1}$) is $\sim 60\times 10^{-3}$ N$\cdot$m$^{-1}$. Over the course of an experiment, water replaces CA removed from the boat starting at the tablet's periphery and progressively proceeds radially inwards towards the tablet's center. Consequently, CA concentration at the c-boat edge constantly decreases resulting in weaker interfacial tension gradients which decrease the boat speed as time progresses; this bears upon results to be discussed. In fig.~\ref{fig2}a, we show the mean c-boat speed (running average over 1 minute interval) monotonically decreases with time.

With decreasing marangoni force strength in time, the c-boat exhibits three distinct modes of motion.  Figure~\ref{fig2}b-d shows time traces of instantaneous c-boat speed at specific intervals corresponding to these three modes. At early times and high surface tension difference, the c-boat speed exhibits harmonic oscillations about the mean (see fig.~\ref{fig2}b). This mean speed and oscillation frequency continuously decrease with time, whereas the oscillation amplitude passes through a maximum before it decreases and transitions to the second distinct mode where the c-boat moves with steady speed (see fig.~\ref{fig2}c). The third distinct mode of relaxation oscillations emerges at long times and low surface tension differences where c-boat motion occurs in periodic bursts interspersed with durations of almost no c-boat motion (see fig.~\ref{fig2}d).

\begin{figure*}[ht] 
    \centering
       \includegraphics[width=\textwidth]{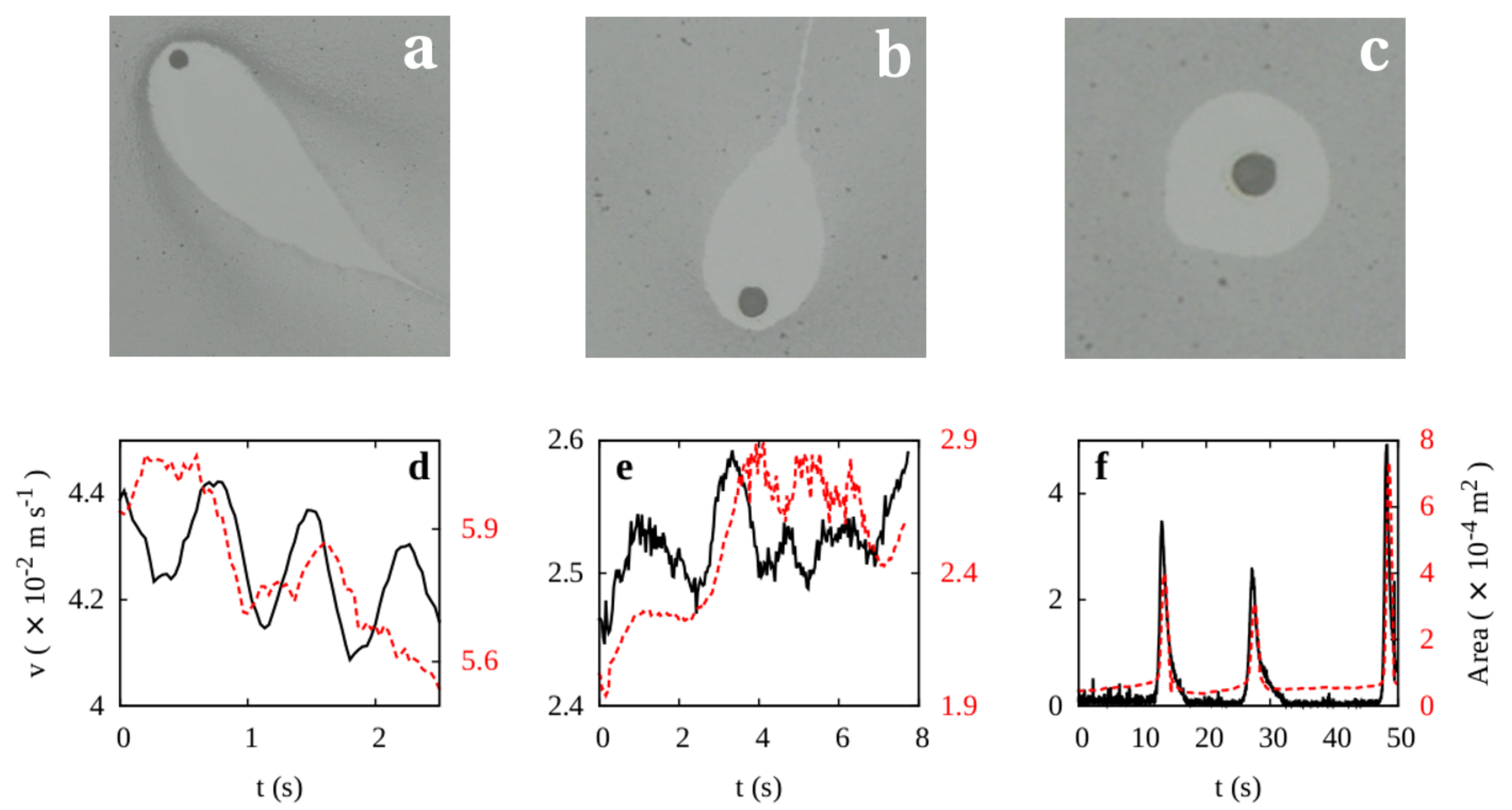}
    \caption{CA spreading from a moving c-boat creates a tracer-free hole (light grey) around a tracer rich (dark grey) region. The qualitative difference observed in tracer free regions for (a) Harmonic: long elliptical comet shape, (b) Steady: short elliptical comet shape with long wispy tail, and (c) Relaxation Oscillation: symmetric circular CA spread (when the c-boat is stationary) can be quantified against c-boat speed as shown for (d) Harmonic, (e) Steady, and (f) Relaxation Oscillation mode, where the CA spread area changes in sync with speed for all three modes. }
    \label{fig3}
\end{figure*}

The evolution of Marangoni force strength with time can be confirmed by visualising the c-boat motion on an interface decorated with buoyant tracer particles. Before proceeding, a few details concerning the behaviour of tracers at air-water interfaces themselves is in order. Tracer particles introduced at the air-water interface form clusters due to inter-particle capillary attraction \cite{Vella2005}. At sufficiently high concentrations, these tracers form a system spanning cluster of jammed particles that occupy the entire interface. A c-boat moving on such a tracer-rich surface exhibits a tracer-free region around the c-boat. The CA spreading from the c-boat onto the interface pushes the tracers away to create the tracer-free hole, and hence forms a visual representation of the spatial extent to which CA acts around a c-boat. The extent of the tracer-free hole is a function of both CA and tracer concentrations. Whereas the spreading CA pushes the tracers away from the c-boat, the tracers provide a back reaction force and attempt to close the hole. This behaviour is true only for hydrophilic particles, and not for hydrophobic tracer particles. Since hydrophilic particles are wettable, they form a thin lubrication film of water on their surfaces, which allows tracer particles pushed against each other to exploit the lubrication film and slide over each other. Once the c-boat passes and CA dissolves in water, the tracers again slide back onto the interface to cover it. On the other hand, hydrophobic particles not being wettable, do not slide over each other, but sustain stresses when pushed against each other to form a two-dimensional solid. A c-boat, if introduced on a hydrophobic tracer-rich interface exerts an external stress on this two-dimensional solid \cite{Vella2004, Vella2006, Mahadevan2011}, which is then relieved {\it via} fracture of the particulate interface \cite{Vella2006, Mahadevan2011}. Whereas both types of tracers provide a back reaction force, they follow from different causes. Indeed, whereas hydrophilic particles behave in a reversible manner, hydrophobic particles exhibit irreversible behaviour through fracture of the two-dimensional particulate interface \cite{Vella2006, Mahadevan2011}. Whereas it is possible to measure this back reaction force through sensitive capacitance measurements \cite{Varshney2012}, they work ideally under quasi-static conditions, and not under the dynamic conditions studied here. Since, we were unable to measure or systematically control this back reaction force, a fixed concentration of hydrophilic buoyant tracers was introduced on the interface prior to start of an experiment, and the same particle concentration was maintained throughout the experiment. This kept the tracer back reaction force fixed for a given experiment; any change in tracer-free area around the c-boat with time therefore followed only from change in CA concentration with time.

\begin{figure*}[ht] 
    \centering
       \includegraphics[width=\textwidth]{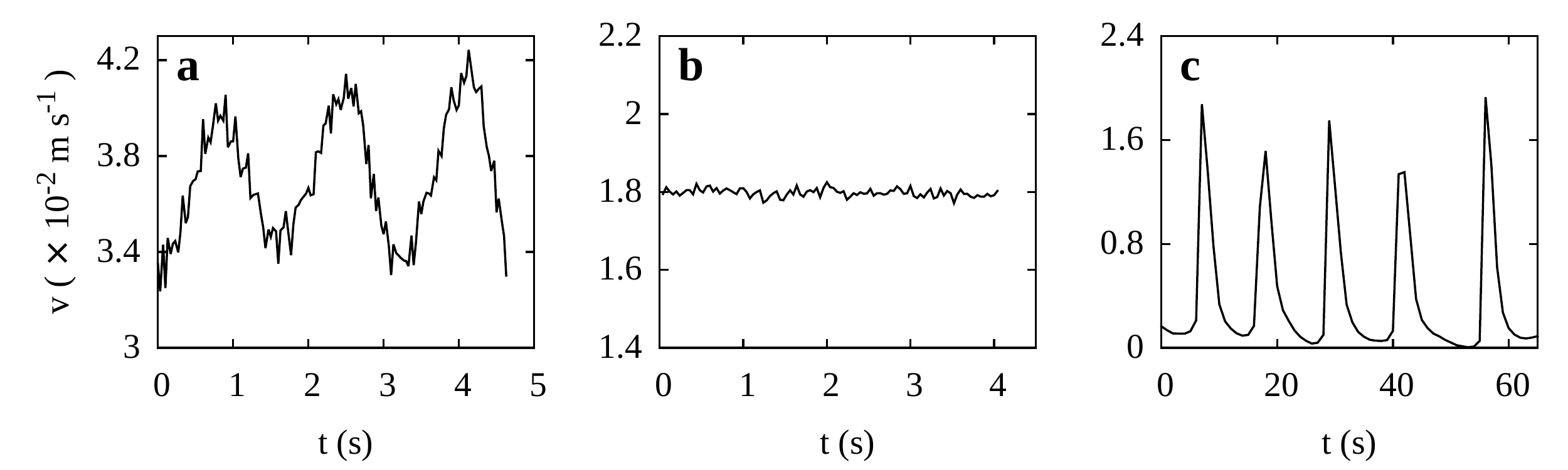}
    \caption{Changing CA-water surface tension difference via metered dosage of SDS in water allows one to demonstrate all three c-boat propulsion modes: (a) Harmonic mode at $\Delta \gamma =  (72\times 10^{-3} - 60\times 10^{-3}) = 12\times 10^{-3}$ N$\cdot$m$^{-1}$, (b) Steady mode at $\Delta \gamma =  (68\times 10^{-3} - 60\times 10^{-3}) = 8\times 10^{-3}$ N$\cdot$m$^{-1}$, and (c) Relaxation Oscillation mode at $\Delta \gamma =  (65\times 10^{-3} - 60\times 10^{-3}) = 5\times 10^{-3}$ N$\cdot$m$^{-1}$}
    \label{fig4}
\end{figure*}

The tracer-free region representing a comet shape also provides a visualization of the asymmetry in the CA distribution around the c-boat. The shape and size of the comet trail is, therefore, indicative of the nature of underlying dynamics. Direct visualization of the comet trail using tracer particles for the three modes of c-boat motion is shown in fig.~\ref{fig3}a-c and the accompanying movies in supplementary information. The reducing area of the comet trail suggests a decrease in the net marangoni force strength propelling the c-boat. The elliptical, comet shaped tracer free region in the harmonic mode (fig.~\ref{fig3}a) gives way to a partially circular comet shape with a thin wispy tail in the steady mode (fig.~\ref{fig3}b). The relaxation oscillation mode results in a stationary c-boat with a circular region that grows with time, until a sudden jump in c-boat position results in a momentary comet shape before the c-boat once again settles into a stationary position with a growing circular area (fig.~\ref{fig3}c). This visualisation can also be quantified by isolating the area of the tracer-free region in each mode and tracking its time evolution, as shown in fig.~\ref{fig3}d-f. Indeed, in the harmonic mode, the area of the tracer-free comet shaped region oscillates in tandem with the sinusoidal oscillation in c-boat speed. This gives way to a more steady variation (modulo noise introduced by ambient fluctuations) in the tracer-free area and c-boat speed in second mode of steady-state motion. However, the most striking confirmation is provided for the third, relaxation oscillation mode (fig.~\ref{fig3}f), where both c-boat speed and the area of the tracer-free region show a sudden spike between long intervals of inactivity.

The steady decrease in mean c-boat speed over the experimental duration (fig.~\ref{fig2}a) points to the steady decrease in Marangoni force due to loss of CA concentration. To further verify that the Marangoni force strength is indeed the parameter governing the transitions from one mode to the other, we independently changed the ambient surface tension. By introducing different amounts of SDS in the bulk water, the surface tension ($\gamma$) was reduced from $\gamma = 72\times 10^{-3}$ N$\cdot$m$^{-1}$ (pure water) down to $\gamma = 59\times 10^{-3}$ N$\cdot$m$^{-1}$. The motion of  freshly prepared c-boats was observed a fixed duration (2 - 3 minutes, depending on experimental run) after the start of experiment in order to allow initial transients to settle down. The c-boat in the presence of SDS exhibited the three modes of motion in the same order as the ambient surface tension was continuously decreased, thereby providing a correspondence between time and surface tension difference. As the time traces of c-boat speed in fig.~\ref{fig4} show, the c-boat motion shows harmonic oscillations for ambient surface tension value of $\gamma = 72\times 10^{-3}$ N$\cdot$m$^{-1}$ (surface tension difference $\Delta \gamma = 12 \times 10^{-3}~\text{N} \cdot \text{m}^{-1}$), steady motion for $\gamma = 68\times 10^{-3}$ N$\cdot$m$^{-1}$ ($\Delta \gamma = 8 \times 10^{-3}~\text{N} \cdot \text{m}^{-1}$), and relaxation oscillations for $\gamma = 65\times 10^{-3}$ N$\cdot$m$^{-1}$ ($\Delta \gamma = 5 \times 10^{-3}~\text{N} \cdot \text{m}^{-1}$). The comet tail structure also shows features similar to those observed when the transitions were a result of CA depletion from the c-boat (fig.~\ref{fig2}). While introduction of SDS may change the Marangoni forces on the interface, our observations suggest that the qualitative behaviour of the c-boat motion is preserved under these dynamics.

\section{Discussion}
\label{discusssec}
A qualitative explanation of the experimental results is possible {\it via} an analogy with electronic oscillators \cite{Diefenderfer1994, Strogatz2001}. Let us suppose, the CA molecules are analogous to electric charge, the surface tension difference acts like a voltage (potential difference), and the air-water interface is akin to a capacitor being charged. The CA molecules spreading on the interface under Marangoni stress are balanced by the shear stress from the underlying fluid, which acts like a resistor controlling the capacitor's charging rate. The capacitor charges until a threshold voltage is attained, at which point it discharges {\it via} a dissipative element (e.g. a bulb). Similarly, the CA molecules populate the interface until they attain a concentration high enough to propel the c-boat, which upon motion is acted upon by the hydrodynamic drag force. The hydrodynamics separately govern the surfactant transport and the drag on the c-boat and thereby determine the character of the charging-discharging dynamics. Since the c-boat neither accelerates nor decelerates over a cycle, the Marangoni force and the hydrodynamic drag must balance each other on average. The observed dynamics result from the magnitude of these forces, and the response-time these forces take to establish, which is governed by separate hydrodynamic processes. {\it Ergo} the charging (CA influx) time and discharge (hydrodynamic drag) represent two time scales $\tau_C$ and $\tau_D$. Both the strength of the Marangoni force and the two time-scales are determined by the surface-tension difference. When the surface tension difference is strong, the Marangoni force driving the c-boat is strong and rapid, but the hydrodynamics responsible for establishing the drag on the c-boat lags. The lag could result from the inertia of the c-boat or the surrounding fluid. This case is underscored by the parameter regime $\tau_C << \tau_D$. The drag lagging behind the driving Marangoni force paves way for the observed harmonic oscillations, i.e. Mode 1 of c-boat propulsion.  As the surface tension difference weakens, along with the strength of the Marangoni force, the charging process slows down. The CA influx onto the surface that drives the c-boat occurs over a time period comparable to the time scale over which the hydrodynamic drag acts on the propelled c-boat, i.e. $\tau_C \simeq \tau_D$. The drag force can keep-up with the driving Marangoni force, and a terminal velocity representing Mode 2 of steady-state c-boat motion is established. Finally, when the surface tension difference decreases enough that the Marangoni force takes a long time to establish compared to the hydrodynamic response ($\tau_C >> \tau_D$),  it naturally leads to the relaxation oscillations represented by Mode 3. This qualitative picture is confirmed by the visualisation of the interface with tracer particles (fig.~\ref{fig3}a-c and supplementary material).

Our experimental results and the above description places constraints on simplified mathematical models capable of a quantitative explanation of the Marangoni-driven propulsion of c-boats. Clearly, such a mathematical model must track the position and velocity, and the unbalanced force on the c-boat as a combination of the hydrodynamic drag and the Marangoni force. An unbalanced Marangoni force results from a combination of the amount of surfactant around the c-boat, its ability to lower the surface tension, and the asymmetry in its surface distribution, as indicated by the comet trails. The mathematical model must allow the asymmetry in the surfactant distribution to result from a spontaneous symmetry breaking that chooses the direction of propagation of the c-boat. Furthermore, successively decreasing the Marangoni force strength in the model must lead to a transition from harmonic to steady to relaxation-oscillation behaviour, in that order. However, a fundamental difference separates the electrical analogy from the hydrodynamic case of c-boat self-propulsion. The resistor controlling the capacitor's charging rate, and hence the time scale $\tau_C$, follows linear dynamics in the electrical analogue by virtue of Ohm's law. For the c-boat dynamics, CA flux exhibits strongly nonlinear and transient transport behaviour as the Marangoni stress is balanced by drag. Indeed, this nonlinear transport relation is not yet known, and warrants a separate investigation. Whereas a detailed model is beyond the scope of this article for reasons described above, here we have presented the key experimental observations that provide the hypothesis, the essential ingredients for its construction, and expected behaviour.

\section{Summary}
In summary, we have studied the self-motion of a camphoric acid boat at the air-water interface and identified three distinct modes of motion. Each of the harmonic, steady-state, and relaxation oscillation modes have been separately reported \cite{Hayashima2001, Suematsu2010, Jin2012, Velev2012} for objects undergoing Marangoni propulsion. We have explained the existence of these propulsive modes as a consequence of CA concentration or Marangoni force strength, which we verified by controlling the surface tension of ambient interface through controlled doses of SDS. We anticipate these modes will contribute to the understanding of collective interactions between multiple c-boats within the context of active matter physics \cite{Ramaswamy2010} as well as autonomous motion and self-assembly \cite{Ismagilov2002}, the hydrodynamics of spreading of camphor acid and other volatile surfactants at air-water interfaces \cite{Troian1998} as well as viscous marangoni propulsion \cite{Lauga2012} among other studies.

\begin{acknowledgement}
VSA, DKS, and MMB were supported by the Collective Interactions Unit, OIST. MMB acnkowledges L. Mahadevan for introducing the camphor boat system and scientific discussions, and D. Vu Anh for help with preliminary experiments. The authors acknowledge Kenneth Meacham III for experimental support, Prof. Amy Shen for help with tensiometry measurements, and helpful discussions with Prof. Pinaki Chakraborty.
\end{acknowledgement}

% \begin{suppinfo}

% This will usually read something like: ``Experimental procedures and
% characterization data for all new compounds. The class will
% automatically add a sentence pointing to the information on-line:

% \end{suppinfo}

% The appropriate \bibliography command should be placed here.
% Notice that the class file automatically sets \bibliographystyle
% and also names the section correctly.

\providecommand{\latin}[1]{#1}
\providecommand*\mcitethebibliography{\thebibliography}
\csname @ifundefined\endcsname{endmcitethebibliography}
  {\let\endmcitethebibliography\endthebibliography}{}

% \bibliography{dcboat}

% \newpage
% \begin{figure*}
% \centerline{\includegraphics[width=8.5cm]{figures/figtoc}}
% \caption{For table of contents use only.}
% \end{figure*}

\end{document}